\documentclass[aps,prl,twocolumn,notitlepage,superscriptaddress]{revtex4-1}

\usepackage{amsmath}
\usepackage{amssymb}
\usepackage{graphicx}
\usepackage{epstopdf}
\usepackage[colorlinks=true]{hyperref}

\begin{document}
\title{Self-Focusing Skyrmion Racetracks in Ferrimagnets}

\author{Se Kwon Kim}
\affiliation{Department of Physics and Astronomy, University of California, Los Angeles, California 90095, USA}
\author{Kyung-Jin Lee}
\affiliation{Department of Materials Science and Engineering, Korea University, Seoul 02841, South Korea}
\affiliation{KU-KIST Graduate School of Converging Science and Technology, Korea University, Seoul 02841, South Korea}
\author{Yaroslav Tserkovnyak}
\affiliation{Department of Physics and Astronomy, University of California, Los Angeles, California 90095, USA}

\date{\today}

\begin{abstract}
We theoretically study the dynamics of ferrimagnetic skyrmions in inhomogeneous metallic films close to the angular momentum compensation point. In particular, it is shown that the line of the vanishing angular momentum can be utilized as a self-focusing racetrack for skyrmions. To that end, we begin by deriving the equations of motion for the dynamics of collinear ferrimagnets in the presence of a charge current. The obtained equations of motion reduce to those of ferromagnets and antiferromagnets at two special limits. In the collective coordinate approach, a skyrmion behaves as a massive charged particle moving in a viscous medium subjected to a magnetic field. Analogous to the snake orbits of electrons in a nonuniform magnetic field, we show that a ferrimagnet with the nonuniform angular momentum density can exhibit snake trajectories of skyrmions, which can be utilized as racetracks for skyrmions.
\end{abstract}

\maketitle

\emph{Introduction.}|A free particle with the magnetic moment precesses at the frequency proportional to the applied magnetic field and its gyromagnetic ratio, which is the ratio of its magnetic moment to its angular momentum. When a magnet is composed of equivalent atoms, its net magnetization and net angular momentum density are collinear with the proportionality given by the gyromagnetic ratio of constituent atoms. Due to the linear relationship between them, the magnetization and the angular momentum density represent the same degrees of freedom, and thus are interchangeable in describing the magnetization dynamics. One-sublattice ferromagnets and two-sublattice antiferromagnets are examples of such magnets. 

When a magnet consists of inequivalent atoms, however, its net magnetization and net angular momentum density can be independent degrees of freedom~\cite{WangsnessPR1953, *WangsnessPR1954, *WangsnessAJP1956}. One class of such magnets is rare-earth transition-metal (RE-TM) ferrimagnetic alloys~\cite{*[][{, and references therein.}] KirilyukRPP2013}, in which the moments of TM elements and RE elements tend to be antiparallel due to the exchange interaction. Because of different gyromagnetic ratios between RE and TM elements, one can reach the angular momentum compensation point and the magnetization compensation point, by varying the relative concentrations of the two species or changing the temperature. These compensation points are absent in the ferromagnets and antiferromagnets, which have been mainstream materials in spintronics~\cite{ZuticRMP2004, *JungwirthNN2016}, and thus may bring a novel phenomenon to the field. In particular, we would like to focus on the dynamics of ferrimagnets around the angular momentum compensation point in this Letter for the following reason. Away from the compensation point, the dynamics of ferrimagnets is close to that of ferromagnets~\cite{AndreevSPU1980}. At the compensation point, its dynamics is antiferromagnetic~\cite{KirilyukRPP2013, KimNM2017}. Therefore, the ideal place to look for the unique aspects of the dynamics of ferrimagnets would be close to, but not exactly at, the angular momentum compensation point. 

Topological solitons in magnets~\cite{*[][{, and references therein.}] KosevichPR1990} have been serving as active units in spintronics. For example, a domain wall, which is a topological soliton in quasi-one-dimensional magnets with easy-axis anisotropy, can function as a memory unit, as demonstrated in the magnetic domain-wall racetrack memory~\cite{ParkinScience2008}. Two-dimensional magnets with certain spin-orbit coupling can also stabilize another particle-like topological soliton, which is referred to as a skyrmion. Skyrmions have been gaining attention in spintronics as information carriers, alternative to domain walls, because of fundamental interest as well as their practical advantages such as a low depinning electric current~\cite{*[][{, and references therein.}] NagaosaNN2013}. Several RE-TM thin films such as GdFeCo and CoTb have been reported to possess the perpendicular magnetic anisotropy and the bulk Dzyaloshinskii-Moriya interaction~\cite{TonoAPE2015, FinleyPRA2016}, and thus are expected to be able to host skyrmions under appropriate conditions. 

In this Letter, we study the dynamics of skyrmions in metallic collinear ferrimagnets, with a specific goal to understand and utilize the dynamics of skyrmions close to the angular momentum compensation point in RE-TM alloys. To that end, we first derive the equations of motion for the dynamics of general collinear magnets in the presence of an electric current. The resultant equations of motion reduce to those of ferromagnets and antiferromagnets at two limiting cases. The dynamics of a skyrmion is then derived within the collective coordinate approach~\cite{TretiakovPRL2008, *TvetenPRL2013}. Generally, it behaves as a massive charged particle in a magnetic field moving in a viscous medium. When there is a line in the sample across which the net angular momentum density reverses its direction, the emergent magnetic field acting on skyrmions also changes its sign across it. Motivated by the existence of a narrow channel in two-dimensional electron gas localized on the line across which the perpendicular magnetic field changes its direction~\cite{MullerPRL1992, *ReijniersJPCM2000}, we show that, under suitable conditions, the line of the vanishing angular momentum in RE-TM alloys can serve as a self-focusing racetrack for skyrmions~\cite{FertNN2013} as a result of combined effects of the effective Lorentz force and the viscous force. We envision that ferrimagnets with the tunable spin density can serve as a natural platform to engineer an inhomogeneous emergent magnetic field for skyrmions, which would provide us a useful knob to control them.

\emph{Main results.}|The system of interest to us is a two-dimensional collinear ferrimagnet. Although the angular momentum can be rooted in either the spin or the orbital degrees of freedom, we will use the term, spin, as a synonym of angular momentum throughout for the sake of brevity. For temperatures much below than the magnetic ordering temperature, $T \ll T_c$, the low-energy dynamics of the collinear ferrimagnet can be described by the dynamics of a single three-dimensional unit vector $\mathbf{n}$, which determines the collinear structure of the magnet~\cite{AndreevSPU1980}. Our first main result, which will be derived later within the Lagrangian formalism taken by \textcite{AndreevSPU1980} for the magnetic dynamics in conjunction with the phenomenological treatment of the charge-induced torques~\cite{HalsPRL2011}, is the equations of motion for the dynamics of $\mathbf{n}$ in the presence of a charge current density $\mathbf{J}$ and an external field $\mathbf{h}$ to the linear order in the out-of-equilibrium deviations $\dot{\mathbf{n}}$, $\mathbf{J}$, and $\mathbf{h}$:
\begin{equation}
\label{eq:main1}
\begin{split}
s \dot{\mathbf{n}} + s_\alpha \mathbf{n} \times \dot{\mathbf{n}} + \rho \mathbf{n} \times \ddot{\mathbf{n}} = & \mathbf{n} \times \mathbf{f}_n + \xi (\mathbf{J} \cdot \boldsymbol{\nabla}) \mathbf{n} \\
& + \zeta \mathbf{n} \times (\mathbf{J} \cdot \boldsymbol{\nabla}) \mathbf{n} \, ,
\end{split}
\end{equation}
where $s$ is the net spin density along the direction of $\mathbf{n}$, $s_\alpha$ and $\rho$ parametrize the dissipation power density $P = s_\alpha \dot{\mathbf{n}}^2$ and the inertia associated with the dynamics of $\mathbf{n}$, respectively, and $\mathbf{f}_n \equiv - \delta U / \delta \mathbf{n}$ is the effective field conjugate to $\mathbf{n}$ with $U[\mathbf{n}]$ the potential energy~\footnote{In the supplemental material, the equations of motion for the dynamics of $\mathbf{n}$ are derived more microscopically for two-sublattice collinear ferrimagnets, which provides us a concrete example of more general cases discussed in the main text.}. Here, $\xi$ and $\zeta$ are the phenomenological parameters for the adiabatic and nonadiabatic torques due to the current, respectively. It is instructive to interpret $\xi \mathbf{J}$ as the product of the dimensionless factor $\tilde{\xi} \equiv \xi / (\hbar / 2 e)$ and the spin current density corresponding to the charge current density, $\mathbf{J}_s \equiv (\hbar / 2e) \mathbf{J}$, where $e < 0$ is the electric charge of conducting electrons. Hereafter, the symbols with the tilde will denote the dimensionless quantities. 

When the inertia vanishes, $\rho = 0$, the obtained equations of motion is reduced to the Landau-Lifshitz-Gilbert equation for ferromagnets augmented by the spin-transfer torques~\cite{SlonczewskiJMMM1996, *BergerPRB1996, ZhangPRL2004, *ThiavilleEPL2005}, in which $s_\alpha / s$ and $\tilde{\xi}$ can be identified as the Gilbert damping constant and the spin polarization rate of conducting electrons, respectively. When the net spin density vanishes, $s = 0$, it corresponds to the equations of motion for antiferromagnets~\cite{HalsPRL2011}. The equations of motion for the dynamics of a two-sublattice ferrimagnet in the absence of an electric current and dissipation, $s_\alpha = 0$ and $\mathbf{J} = 0$, has been obtained by \textcite{IvanovJETP1983}.

The low-energy dynamics of rigid magnetic solitons in two-dimensional collinear magnets can be derived from Eq.~(\ref{eq:main1}) within the collective coordinate approach~\cite{TretiakovPRL2008}, where the dynamics of the order parameter is encoded in the time evolution of the soliton position, $\mathbf{n} (\mathbf{r}, t) = \mathbf{n}_0 [\mathbf{r} - \mathbf{R} (t)]$. The resultant equations of motion for the position of a circularly symmetric soliton, which are obtained by integrating Eq.~(\ref{eq:main1}) multiplied by $\mathbf{n}_0 \times \partial_\mathbf{R} \mathbf{n}_0$ over the space, are our second main result:
\begin{equation}
\label{eq:main2}
M \ddot{\mathbf{R}} = Q \dot{\mathbf{R}} \times \mathbf{B} - D \dot{\mathbf{R}} + \mathbf{F}_U + \mathbf{F}_J \, ,
\end{equation}
where $M \equiv \rho \int dxdy (\partial_x \mathbf{n}_0)^2$ is the soliton mass~\footnote{Relaxation of the rigidity approximation for the soliton structure will give rise to additional contributions to its mass from the internal fast modes~\cite{MakhfudzPRL2012}. Therefore, understanding the dynamics of general solitons would require us to consider the mass $M$ as a parameter that can be different from the given expression.}, $D \equiv s_\alpha \int dxdy (\partial_x \mathbf{n}_0)^2$ is the viscous coefficient, $\mathbf{F}_U \equiv - d U / d \mathbf{R}$ is the internal force, 
$
(F_J)_i \equiv \int dxdy [ \xi \mathbf{n} \cdot (\mathbf{J} \cdot \boldsymbol{\nabla}) \mathbf{n} \times \partial_i \mathbf{n} - \zeta \partial_i \mathbf{n} \cdot (\mathbf{J} \cdot \boldsymbol{\nabla}) \mathbf{n} ]
$
is the force due to the charge current. The first term on the right-hand side is the effective Lorentz force on the soliton, which is proportional to its topological charge
\begin{equation}
Q = \frac{1}{4 \pi} \int dxdy \, \mathbf{n}_0 \cdot (\partial_x \mathbf{n}_0 \times \partial_y \mathbf{n}_0) \, ,
\end{equation}
which measures how many times the unit vector $\mathbf{n}_0 (\mathbf{r})$ wraps the unit sphere as $\mathbf{r}$ spatially varies~\cite{BelavinJETP1975}, and the fictitious magnetic field
\begin{equation}
\mathbf{B} \equiv B \hat{\mathbf{z}} = - 4 \pi s \hat{\mathbf{z}} \, .
\end{equation}

According to the equations of motion, a skyrmion in chiral ferrimagnets, which is characterized by its topological charge $Q = \pm 1$, behaves as a massive charged particle in a magnetic field moving in a viscous medium. The fictitious magnetic field is proportional to the net spin density $s$ along the direction of the order parameter $\mathbf{n}$, which leads us to consider collinear magnets with tunable $s$ to look for a possibly interesting dynamics of a skyrmion. The RE-TM ferrimagnetic alloys~\cite{KirilyukRPP2013} are such materials. For example, Co$_{1-x}$Tb$_x$ has been shown to exhibit the vanishing angular momentum $s \approx 0$ at $x \approx 17$\% at room temperature~\cite{FinleyPRA2016} by varying the chemical composition. As another example, the angular momentum compensation temperature of Gd$_{22\%}$Fe$_{75\%}$Co$_{3\%}$ has been reported as $T \approx 220$K~\cite{StanciuPRB2006}.

\begin{figure}
\includegraphics[width=\columnwidth]{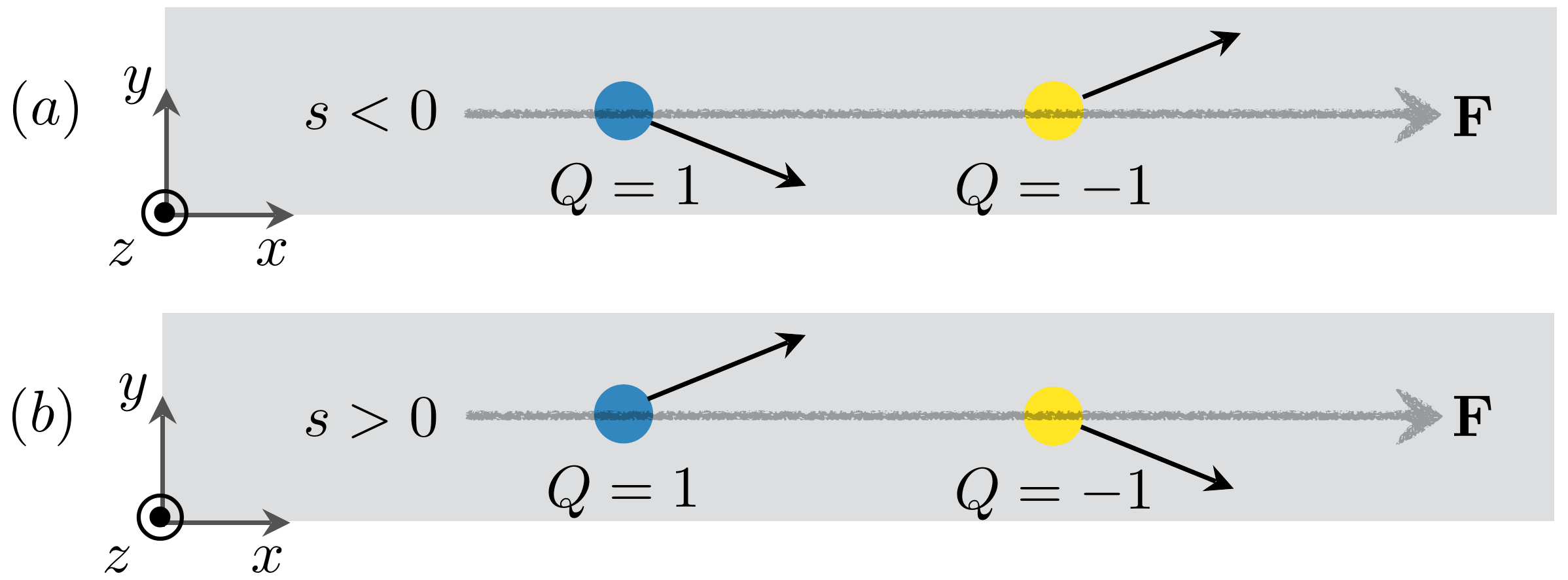}
\caption{Schematic illustrations of a steady-state skyrmion motion [Eq.~(\ref{eq:V})] in the presence of a current-induced force $\mathbf{F} = F \hat{\mathbf{x}}$. Four possible types are classified by its skyrmion charge $Q$ and the sign of the net spin density $s$. See the main text for the discussions.}
\label{fig:fig1}
\end{figure}

A skyrmion can be driven by an electric current as can be seen in Eq.~(\ref{eq:main2}). In the presence of the corresponding current-induced force $\mathbf{F}_J \equiv F \hat{\mathbf{x}}$, the direction of which is defined as the $x$ axis, the steady state of a skyrmion is given by
\begin{equation}
\label{eq:V}
\dot{\mathbf{R}} \rightarrow \mathbf{V} = \frac{F}{B^2 + D^2} \left( D \hat{\mathbf{x}} - Q B \hat{\mathbf{y}} \right) \, .
\end{equation}
See Fig.~\ref{fig:fig1} for illustrations of a steady-state skyrmion motion for $F > 0$. The skyrmion with the topological charge $Q = 1$ moves down for $s < 0$ and up for $s > 0$, while moving to the right regardless of the sign of $s$. If the ferrimagnet is prepared in such a way that $s < 0$ for $y > 0$ and $s > 0$ for $y < 0$, the skyrmion with $Q = 1$ will move along the horizontal line $y = 0$ after certain relaxation time because it is constantly pushed back to the line via the effective Lorentz force. Note that the skyrmion experiences no Lorentz force on the angular momentum compensation line, and thus will move as an antiferromagnetic skyrmion along it~\cite{BarkerPRL2016, *ZhangSR2016}.

\begin{figure}
\includegraphics[width=\columnwidth]{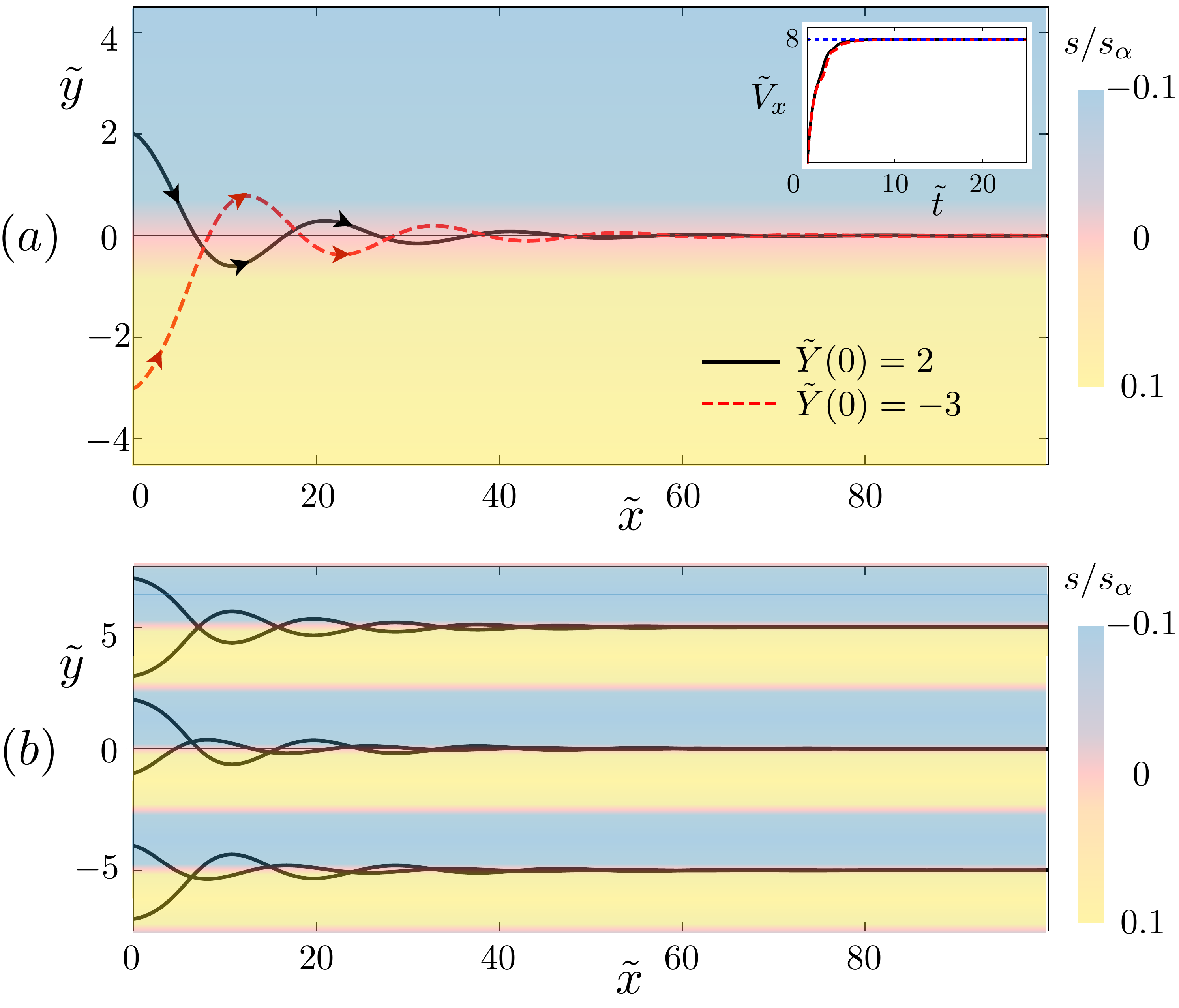}
\caption{Trajectories of skyrmions with the topological charge $Q = 1$ in the presence of a current-induced force $\mathbf{F} =F \hat{\mathbf{x}}$, which are obtained by numerically solving the dimensionless equations of motion for the dynamics of skyrmions in Eq.~(\ref{eq:numeric}). (a) Two trajectories for the monotonic net angular momentum density $s$. The inset shows the convergence of the skyrmion velocities. (b) Multiple trajectories for the periodic net angular momentum density $s$. See the main text for the detailed discussions.}
\label{fig:fig2}
\end{figure}

To corroborate the qualitative prediction, we numerically solve the equations of motion [Eq.~(\ref{eq:main2})] in its dimensionless form:
\begin{equation}
\label{eq:numeric}
I \frac{d^2 \tilde{\mathbf{R}}}{d \tilde{t}^2} + \frac{4 \pi s Q}{s_\alpha} \frac{d \tilde{\mathbf{R}}}{d \tilde{t}} \times \hat{\mathbf{z}} + I \frac{d \tilde{\mathbf{R}}}{d \tilde{t}} = \tilde{F} \hat{\mathbf{x}} \, ,
\end{equation}
in which time, length, and energy are measured in units of the relaxation time $\tau \equiv \rho / s_\alpha$, the characteristic length scale for the skyrmion size $l$~\footnote{For example, the energy density, $\mathcal{U} = A (\boldsymbol{\nabla} \mathbf{n})^2 / 2 - K n_z^2 / 2 + D \mathbf{n} \cdot (\boldsymbol{\nabla} \times \mathbf{n})$, yields the characteristic length scale for the skyrmion radius, $l = D / K$~\cite{BogdanovJETP1989}.}, and $\epsilon \equiv s_\alpha^2 l^2 / \rho$, respectively, where $I = \int dxdy (\partial_x \mathbf{n}_0)^2$ is a dimensionless number determined by the skyrmion structure. Figure~\ref{fig:fig2}(a) shows the two trajectories of skyrmions of the charge $Q = 1$ departing from $(\tilde{X}, \tilde{Y}) = (0, 2)$ and $(\tilde{X}, \tilde{Y}) = (0, -3)$ with the zero initial velocity under the following configurations: $I = \pi / 2$, $\tilde{F} = 4 \pi$, and $s / s_\alpha = - 0.1 \tanh(\tilde{y}) $. We refer the paths as skyrmion snake trajectories due to their shapes, analogous to the electronic snake orbits in an inhomogeneous magnetic field~\cite{MullerPRL1992}. The inset shows that the skyrmion speed converges as $\tilde{V}_y \rightarrow \tilde{F} / I$ after sufficiently long time, $\tilde{t} \gg 1$. Figure~\ref{fig:fig2}(b) depicts multiple trajectories of skyrmions when the net spin density is spatially periodic, $s / s_\alpha = - 0.1 \sin (2 \pi \tilde{y} / 5)$. Skyrmions are attracted to the angular momentum compensation lines and their velocities converge to the finite value. This leads us to state our third main result: self-focusing narrow guides for skyrmions can be realized in certain ferrimagnets such as the RE-TM alloys along the lines of the angular momentum compensation points, which can be useful in using skyrmions for information processing by, e.g., providing multiple parallel skyrmion racetracks in one sample~\cite{TomaselloSR2014}.

\emph{The dynamics of collinear magnets.}|The derivation of the equations of motion for the dynamics of collinear magnets in [Eq.~(\ref{eq:main1})] is given below, which follows the phenomenological approach taken for antiferromagnets by \textcite{AndreevSPU1980}. Within the exchange approximation that the Lagrangian is assumed invariant under the global spin rotations, we can write the Lagrangian density for the dynamics of the directional order parameter $\mathbf{n}$ in the absence of an external field as
\begin{equation}
\mathcal{L} = - s \mathbf{a}[\mathbf{n}] \cdot \dot{\mathbf{n}} + \frac{\rho \dot{\mathbf{n}}^2}{2} - \mathcal{U} [\mathbf{n}] \, ,
\end{equation}
to the quadratic order in the time derivative, where $\mathbf{a}[\mathbf{n}]$ is the vector potential for the magnetic monopole, $\boldsymbol{\nabla}_{\mathbf{n}} \times \mathbf{a} = \mathbf{n}$~\cite{IvanovSSC1984, *LossPRL1992}. The first term accounts for the spin Berry phase associated with the net spin density along $\mathbf{n}$; The second term accounts for the inertia for the dynamics of $\mathbf{n}$, which can arise due to, e.g., the relative canting of the sublattice spins~\cite{Note1}.

Next, the effects of an external field can be taken into account as follows. The conserved Noether charge associated with the symmetry of the Lagrangian under the global spin rotations is the net spin density, and it is given by $\mathbf{s} = s \mathbf{n} + \rho \mathbf{n} \times \dot{\mathbf{n}}$. The magnetization in the presence of an external field $\mathbf{H}$ can be then written as $\mathbf{M} = g_l s \mathbf{n} + g_t \rho \mathbf{n} \times \dot{\mathbf{n}} + \chi \mathbf{H}$, where $g_l$ and $g_t$ are the gyromagnetic ratios for the longitudinal and transverse components of the spin density with respect to the direction $\mathbf{n}$, respectively, and $\chi$ is the magnetic susceptibility tensor. The relation, $\mathbf{M} = \partial \mathcal{L} / \partial \mathbf{H}$~\cite{AndreevSPU1980}, requires the susceptibility to be $\chi_{ij} = \rho g_t^2 (1 - n_i n_j)$, with which the Lagrangian is extended to
\begin{equation}
\mathcal{L} = - s \mathbf{a}[\mathbf{n}] \cdot \dot{\mathbf{n}} + \frac{\rho (\dot{\mathbf{n}} - g_t \mathbf{n} \times \mathbf{H})^2}{2} - \mathcal{U} [\mathbf{n}] \, ,
\end{equation}
where $\mathcal{U} [\mathbf{n}]$ includes the Zeeman term, $- g_l s \mathbf{n} \cdot \mathbf{H}$. Finally, the dissipation can be accounted for by the Rayleigh dissipation function, $\mathcal{R} = s_\alpha \dot{\mathbf{n}}^2 / 2$, which is the half of the dissipation rate of the energy density, $\mathcal{P} = 2 \mathcal{R}$. The equations of motion obtained from the Lagrangian and the Rayleigh dissipation function are given by Eq.~(\ref{eq:main1}) without the current-induced torques.

\emph{Current-induced torques.}|To derive the torque terms due to an electric current, it is convenient to begin by phenomenologically constructing the expression for the charge current density $\mathbf{J}^\text{pump}$ induced by the magnetic dynamics, and subsequently to invoke the Onsager's reciprocity to obtain the torque terms as done for antiferromagnets in Ref.~\cite{HalsPRL2011}. To the lowest order of the space-time gradients and to the first order in the deviations from the equilibrium, we can write two pumping terms that satisfy the appropriate spatial and spin-rotational symmetries: $\dot{\mathbf{n}} \cdot \partial_i \mathbf{n}$ and $\mathbf{n} \cdot (\dot{\mathbf{n}} \times \partial_i \mathbf{n})$. The resultant expression for the induced current density is given by
\begin{equation}
J^\text{pump}_i / \sigma = \zeta \dot{\mathbf{n}} \cdot \partial_i \mathbf{n} + \xi \mathbf{n} \cdot (\partial_i \mathbf{n} \times \dot{\mathbf{n}}) \, ,
\end{equation}
where $\sigma$ is the conductivity. 

To invoke the Onsager reciprocity that is formulated in the linear order in the time derivative of the dynamic variables, we turn to the Hamiltonian formalism instead of the Lagrangian formalism. We shall restrict ourselves here to the case of a vanishing external field for simplicity, but it can be easily generalized to the case of a finite external field. The canonical conjugate momenta of $\mathbf{n}$ is given by $\mathbf{p} \equiv \partial \mathcal{L} / \partial \dot{\mathbf{n}} = \rho (\dot{\mathbf{n}} - g_t \mathbf{n} \times \mathbf{h}) - s \mathbf{a}$. The Hamiltonian density is then given by
\begin{equation}
\mathcal{H}[\mathbf{n}, \mathbf{p}] = \mathbf{p} \cdot \dot{\mathbf{n}} - \mathcal{L} = \frac{(\mathbf{p} + s \mathbf{a})^2}{2 \rho} + \mathcal{U} \, ,
\end{equation}
which resembles the Hamiltonian for a charged particle subjected to an external magnetic field~\cite{Goldstein}. The Hamilton equations are given by 
\begin{eqnarray}
\dot{\mathbf{n}} &=& \frac{\partial \mathcal{H}}{\partial \mathbf{p}} \equiv - \mathbf{h}_p \, , \\
\dot{\mathbf{p}} &=& - \frac{\partial \mathcal{H}}{\partial \mathbf{n}} - \frac{\partial \mathcal{R}}{\partial \dot{\mathbf{n}}} \equiv \mathbf{h}_n - s_\alpha \dot{\mathbf{n}} = \mathbf{h}_n + s_\alpha \mathbf{h}_p \, ,
\end{eqnarray}
where $\mathbf{h}_p$ and $\mathbf{h}_n$ are conjugate fields to $\mathbf{p}$ and $\mathbf{n}$, respectively. In terms of the conjugate fields, the pumped charge current is given by $\mathbf{J}^\text{pump} = - \zeta \partial_i \mathbf{n} \cdot \mathbf{h}_p - \xi (\mathbf{n} \times \partial_i \mathbf{n}) \cdot \mathbf{h}_p$. By using the Onsager reciprocity and Ohm's law for the current $\mathbf{J} = \sigma \mathbf{E}$, we can obtain the torque terms in Eq.~(\ref{eq:main1}).

\emph{Discussion.}|Let us discuss approximations that have been used in the Letter. First, we have developed the theory for the dynamics of collinear magnets within the exchange approximation~\cite{AndreevSPU1980}, in which the total energy is invariant under the simultaneous rotation of the constituent spins. The relativistic interactions including the magnetic anisotropy, which weakly break the exchange symmetry of the magnet, are added phenomenologically to the potential energy. Secondly, when studying the dynamics of skyrmions in inhomogeneous ferrimagnetic films, we have considered the nonuniform spin density $s$, while neglecting possible spatial variations of the other parameters such as the inertia $\rho$ or the damping $s_\alpha$ because we do not expect those variations to change the results qualitatively. As long as skyrmions are attracted to the line of vanishing angular momentum due to the combined effects of the effective Lorentz force, the viscous force, and the current-induced force, the line should be able to convey skyrmions along with it.

Ferrimagnetic RE-TM alloys have not only the angular momentum compensation point, which we have focused on in this Letter, but also the magnetic moment compensation point. Motivated by the attraction of skyrmions toward the angular momentum compensation lines that we have discussed, it would be worth looking for an interesting phenomenon that can occur on the magnetic moment compensation line. For example, since the magnetic moment governs the magnetostatic energy, there may be unusual magnetostatic spin-wave modes~\cite{DamonJPCS1961, *DamonJAP1965} localized at the line. In addition, we have considered the dynamics of a soliton in two-dimensional ferrimagnets driven by an electric current. In general, the dynamics of a soliton can be induced by other stimuli such as an external magnetic field~\cite{SchryerJAP1974} and a spin-wave excitation~\cite{HinzkePRL2011, *YanPRL2011, *KovalevEPL2012, TvetenPRL2014, KimPRB2014}, which may exhibit peculiar features of ferrimagnets that are absent in ferromagnets and antiferromagnets.

\begin{acknowledgments}
This work was supported by the Army Research Office under Contract No. W911NF-14-1-0016 (S.K.K. and Y.T.) and by the National Research Foundation of Korea (NRF) grant funded by the Korea government (MSIP) (2015M3D1A1070465) (K.-J.L.).
\end{acknowledgments}

\bibliography{/Users/evol/Dropbox/School/Research/master}

\renewcommand{\theequation}{S\arabic{equation}}
\renewcommand{\thefigure}{S\arabic{figure}}
\setcounter{equation}{0}
\setcounter{figure}{0}

\clearpage
\newpage
\appendix
\section{Supplemental Material}

In this supplemental material, we derive the equations of motion for a two-sublattice ferrimagnet by following the approach taken in Ref.~\cite{HalsPRL2011} with the explicit treatment of two sublattices.

\begin{figure}
\includegraphics[width=\columnwidth]{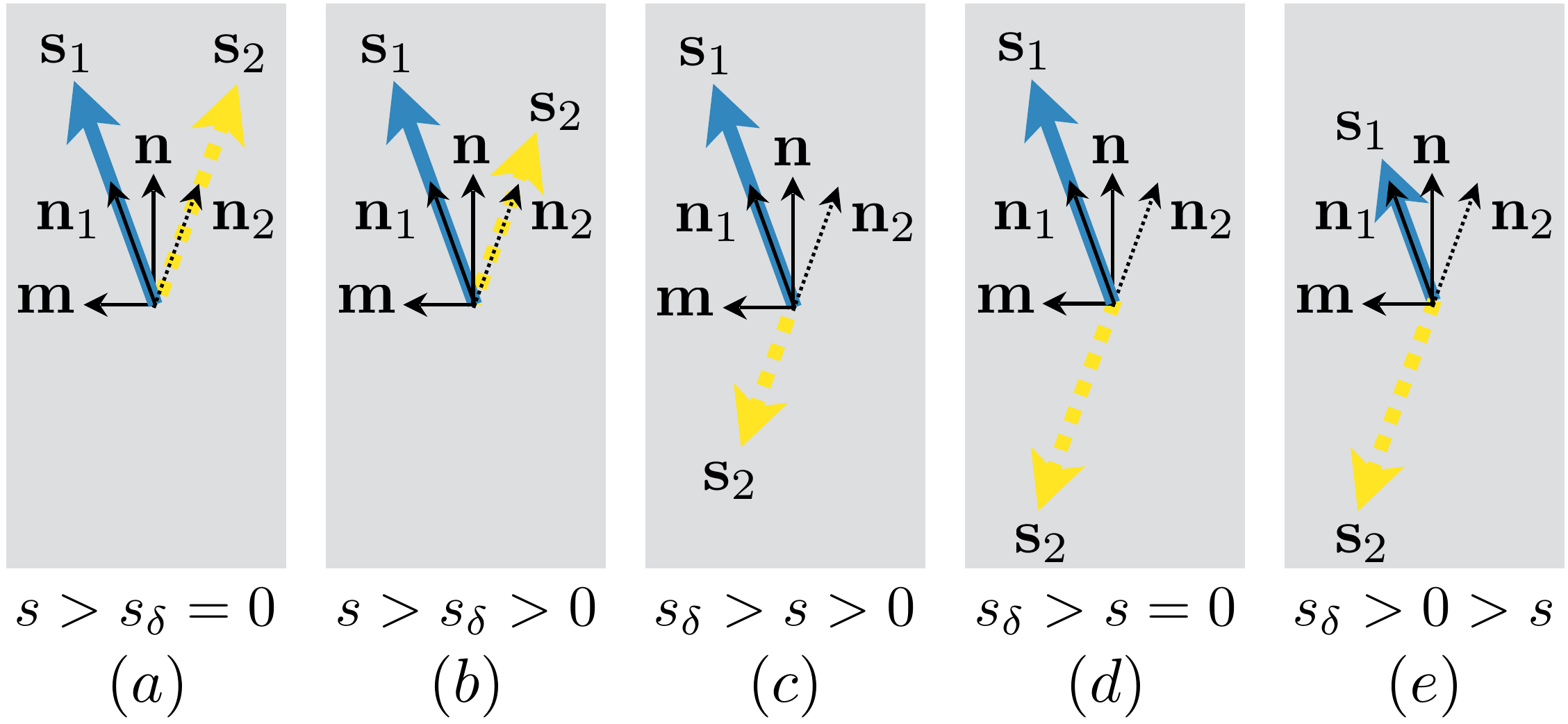}
\caption{Schematic illustrations of possible configurations of the spin densities $\mathbf{s}_1 \equiv s_1 \mathbf{n}_1$ and $\mathbf{s}_2 \equiv s_2 \mathbf{n}_2$ of the two sublattices in collinear magnets, which are classified by the relative magnitude and the sign of the net scalar spin density $s = s_1 + s_2$ and the staggered scalar spin density $s_\delta = (s_1 - s_2) / 2$. (a) and (d) correspond to a one-sublattice ferromagnet and a two-sublattice antiferromagnet, respectively; (b) corresponds to a ferrimagnet, in which the two inequivalent sublattices are coupled by a ferromagnetic exchange; (c) and (e) correspond to ferrimagnets, in which the two inequivalent sublattices are coupled by an antiferromagnetic exchange.}
\label{fig:figs1}
\end{figure}

The model system is a two-dimensional collinear magnet that consists of two inequivalent sublattices. The local spin densities of the two sublattices are denoted by $\mathbf{s}_1 \equiv s_1 \mathbf{n}_1$ and $\mathbf{s}_2 \equiv s_2 \mathbf{n}_2$, where $\mathbf{n}_1$ and $\mathbf{n}_2$ are slowly varying unit vectors. We allow the two scalar spin densities, $s_1$ and $s_2$, to be either positive and negative, which is useful to construct a general theory for collinear magnets as will be shown below. In equilibrium, the two spin densities are collinear, which we represent by $\mathbf{n}_1 = \mathbf{n}_2$. To describe the dynamics of the magnet, it is convenient to use the new vectors, $\mathbf{n} \equiv (\mathbf{n}_1 + \mathbf{n}_2) / 2$ and $\mathbf{m} \equiv \mathbf{n}_1 - \mathbf{n}_2$, instead of $\mathbf{n}_1$ and $\mathbf{n}_2$, and the new scalars, $s = s_1 + s_2$ and $s_\delta = (s_1 - s_2) / 2$, instead of $s_1$ and $s_2$. Here, $\mathbf{n}$ serves as the order parameter, which captures the collinear structure in equilibrium; $\mathbf{m}$ corresponds to the relative canting of the two sublattices, which vanishes in equilibrium; $s$ and $s_\delta$ are the net and the staggered spin densities in equilibrium, respectively. The cases where the two sublattices are coupled by a ferromagnetic exchange can be represented by $s_1,  s_2 > 0$, for which $\mathbf{n}$ is the direction of the net spin density in equilibrium. The cases of an antiferromagnetic exchange can be represented by $s_1 > 0 > s_2$, for which $\mathbf{n}$ is the direction of the staggered spin density in equilibrium. From the definitions, we obtain $\mathbf{n} \cdot \mathbf{m} = 0$, and, for small deviations from the equilibrium, we can impose the constraints $|\mathbf{n}| = 1$ and $|\mathbf{m}| \ll 1$~\cite{HalsPRL2011}. Without loss of generality, we can assume $s_\delta \ge 0$. See Fig.~\ref{fig:figs1} for illustrations of possible types of collinear structures.

Let us first derive the equations of motion in the absence of an electric current within the Lagrangian formalism. The spin Berry phase contribution to the Lagrangian density, which governs the magnetic dynamics, is given by 
\begin{equation}
\mathcal{L}_B = - s_1 \mathbf{a} (\mathbf{n}_1) \cdot \dot{\mathbf{n}}_1 - s_2 \mathbf{a} (\mathbf{n}_2) \cdot \dot{\mathbf{n}}_2 \, ,
\end{equation}
where $\mathbf{a}$ is a vector potential for magnetic monopoles, which satisfies $\boldsymbol{\nabla}_\mathbf{n} \times \mathbf{a}(\mathbf{n}) = \mathbf{n}$. By expanding the spin Berry phase $\mathcal{L}_B$ to the second order in $\mathbf{m}$ and $\dot{\mathbf{n}}$ as done in Ref.~\cite{KimPRB2014}, we obtain
\begin{equation}
\mathcal{L}_B = - s \, \mathbf{a}(\mathbf{n}) \cdot \dot{\mathbf{n}} + s_\delta \, \mathbf{n} \cdot (\dot{\mathbf{n}} \times \mathbf{m}) - s \, \dot{\mathbf{m}} \cdot (\mathbf{n} \times \mathbf{m}) / 8 \, .
\end{equation}
The first term comes from the net spin Berry phase, which is in the Lagrangian for ferromagnets; The second term comes from the cancelation of the spin Berry phases of the two sublattices, which is in the Lagrangian for antiferromagnets; the third term shall be ignored over the first term for the slow dynamics. The total Lagrangian density is given by $\mathcal{L} = \mathcal{L}_B - \mathcal{U}[\mathbf{n}, \mathbf{m}]$. The dissipation can be accounted for by the Rayleigh dissipation function, 
$
\mathcal{R} = s_\alpha \dot{\mathbf{n}}^2 / 2
$,
which is the half of the dissipation rate of the energy density, $\mathcal{P} = 2 \mathcal{R}$. Here, we consider the dissipation associated with the dynamics of the order parameter $\mathbf{n}$, while neglecting the contribution from the dynamics of $\mathbf{m}$ assuming $|\dot{\mathbf{m}}| \ll |\dot{\mathbf{n}}|$. The equations of motion for the fields $\mathbf{n}$ and $\mathbf{m}$ can be obtained from the Lagrangian and the Rayleigh dissipation function~\cite{Goldstein}:
\begin{eqnarray}
s_\delta \dot{\mathbf{n}} &=& \mathbf{n} \times \mathbf{f}_m \, , \\
s_\delta \dot{\mathbf{m}} &=& \mathbf{n} \times (\mathbf{f}_n - s_\alpha \dot{\mathbf{n}}) - (s / s_\delta) \mathbf{n} \times \mathbf{f}_m \, ,
\end{eqnarray}
where $\mathbf{f}_n = - \delta U / \delta \mathbf{n}$ and $\mathbf{f}_m = - \delta U / \delta \mathbf{m}$ are the effective fields conjugate to $\mathbf{n}$ and $\mathbf{m}$, respectively. 

By using the Onsager reciprocity as done in the main text, we can obtain the torque terms in the equations of motion:
\begin{eqnarray}
s_\delta \dot{\mathbf{n}} &=& \mathbf{n} \times \mathbf{f}_m \, , \\
s_\delta \dot{\mathbf{m}} &=& \mathbf{n} \times (\mathbf{f}_n - s_\alpha \dot{\mathbf{n}}) - (s / s_\delta) \mathbf{n} \times \mathbf{f}_m \\ 
&& + \zeta \mathbf{n} \times (\mathbf{J} \cdot \boldsymbol{\nabla}) \mathbf{n} + \xi(\mathbf{J} \cdot \boldsymbol{\nabla}) \mathbf{n} \, . \nonumber
\end{eqnarray}

Within the exchange approximation that the energy is invariant under the global spin rotations, the free energy expanded to the second order in the gradients and the relative canting $\mathbf{m}$ is given by $U[\mathbf{n}, \mathbf{m}] = \int dV [ \mathbf{m}^2 / 2 \chi + A (\partial_i \mathbf{n} \cdot \partial_i \mathbf{n}) / 2 - \mathbf{h} \cdot \mathbf{n} - \mathbf{g} \cdot \mathbf{m} ]$, where $\chi$ represents the magnetic susceptibility, $A$ is the stiffness associated with the spatial change of $\mathbf{n}$, $\mathbf{h} = (M_1 + M_2) \mathbf{H}$, $\mathbf{g} = (M_1 - M_2) \mathbf{H}$, and $\mathbf{H}$ is a static external magnetic field. Here, $M_1 = \gamma_1 s_1$ and $M_2 = \gamma_2 s_2$ are the magnetizations of the two sublattices, where $\gamma_1$ and $\gamma_2$ are their gyromagnetic ratios. Using $\mathbf{f}_m = - \mathbf{m} / \chi + \mathbf{n} \times (\mathbf{g} \times \mathbf{n})$, we can remove $\mathbf{m}$ from the equations of motion, which results in Eq.~(\ref{eq:main1}) with $\rho = s_\delta^2 \chi$.

\end{document}